\documentclass{mem}
\usepackage{natbib}\usepackage{txfonts}\usepackage{balance}
\usepackage{flushend}
\usepackage{graphicx}
\usepackage[breaklinks]{hyperref}
\idline{00}{000}
\begin{document}
\def\teff{$T\rm_{eff }$}
\def\kms{$\mathrm {km s}^{-1}$}

\title{
Towards a joint X-ray and gamma-ray analysis of Pulsar Wind Nebulae with Gammapy
}


\author{
K. \,Egg
\and A. M. W. \, Mitchell
          }

\institute{
Erlangen Centre for Astroparticle Physics (ECAP), Friedrich-Alexander-Universität Erlangen-Nürnberg, Germany
\email{katharina.egg@fau.de}\\
}

\authorrunning{Egg}

\titlerunning{Towards a joint X-ray and gamma-ray analysis of PWNe with Gammapy}

\date{Received: XX-XX-XXXX (Day-Month-Year); Accepted: XX-XX-XXXX (Day-Month-Year)}

\abstract{
For detailed studies of Pulsar Wind Nebulae (PWNe), objects that show photon emission across the entire electromagnetic spectrum, multiwavelength analyses are crucial. The comparison of especially X-ray and gamma-ray emission and their angular sizes can help us to constrain the properties of PWNe, such as their particle transport mechanism or their potential for the acceleration of hadronic particles.
In this vein, we are working towards a joint analysis of eROSITA X-ray data and H.E.S.S. gamma-ray data. To enable this, eROSITA data is adapted into the framework of Gammapy, a Python package for gamma-ray analysis through a multi-step process of adapting the formats of not only the photon event list, but also all X-ray response functions, into open data formats compatible with Gammapy. This is accomplished using newly developed Python converter functions.
In this contribution we present the first eROSITA maps of the PWN MSH 15-52 in Gammapy, compared to the associated H.E.S.S. emission.
}
\maketitle{}

\section{Introduction}
Pulsar wind nebulae (PWNe) are regions of energetic electrons and positrons around pulsars, very interesting sources not only at X-ray and gamma-ray energies but over a wide range of the electromagnetic spectrum. They are created when particles accelerated by a central pulsar interact with the surrounding medium, which causes photon emission \citep{Gaensler_2006}. Among identified sources, PWNe are the most numerous source class in galactic very-high-energy (VHE) gamma-ray astronomy \citep{HGPS_2018}. A number of questions, however, are still open in regards to their properties, such as the nature of their particle acceleration mechanism. Additionally PWNe are candidate PeVatrons, objects that accelerate particles to PeV energies \citep{Mitchell_2022}.

For the study of the extended structures of PWNe in gamma-ray energies the functionalities of the Python package Gammapy are ideal \citep{Donath_2023}. Gammapy has introduced a 3D fitting algorithm, which fits the two spatial dimensions and the energy dimension simultaneously \citep{Mohrmann_2019}. Additionally one of Gammapy's key strengths is its compatibility with multi-mission gamma-ray data. Data from several different types of instruments have been included into its framework, enabling joint analyses over many instruments, such as the flagship Crab spectrum fit by \citet{Nigro_2019}. To further extend the energy range accessible to Gammapy analysis, X-ray 1D (energy) spectra have successfully been introduced into Gammapy by \citet{Giunti_2022} and \citet{Rosillo_2024}.

It is the objective of our work to include X-ray data taken by the eROSITA telescope not only as 1D-spectra, but also at the event level in order to enable detailed spectral as well as morphological analysis through 3D fits, culminating in a combined analysis of X-ray and gamma-ray data. These proceedings will showcase our progress towards this goal and illustrate the conversion process of the X-ray data in to a Gammapy-compatible format.

%

\section{Analysed data}

\subsection{eROSITA data}
The eROSITA X-ray telescope on the SRG satellite has conducted four complete all-sky surveys, preceded by a Calibration and Performance Verification (CalPV) phase \citep{Predehl_2021,Merloni_2024}.
The data of the western Galactic hemisphere taken in the CalPV phase were made public in the early data release (EDR).
These data include a pointed observation of MSH 15-52, that was analysed in the context of this work.

The western half of the first eROSITA all-sky survey (eRASS1) meanwhile, was made public on January 31st 2024 comprising the first data release (DR1)
\citep{Merloni_2024}. MSH 15-52 was also observed in the course of eRASS1, making additional data publicly available for our analysis.

eROSITA consists of seven telescope modules (TMs) recording data simultaneously. Of these TMs 5 and 7 were excluded in this analysis as they are affected by an optical light leak \citep{Predehl_2021}. During the EDR data only TMs 1 to 3 were recording data, all of which were used for our analysis.

\subsection{H.E.S.S. data}
The H.E.S.S. public data release includes 20 runs of 2004 observations of MSH 15-52, equalling 9.1 hours of observations of the extended source \citep{hess_public_data}.
The H.E.S.S. data were used in this work not for the purpose of conducting new scientific analysis but to illustrate the great potential for synergy between eROSITA and H.E.S.S. data through simultaneous and combined analysis within one tool.

\section{File conversion scheme}
In this section the process of converting the eROSITA X-ray data to GADF\footnote{\url{https://gamma-astro-data-formats.readthedocs.io/en/latest/}} Gammapy-compatible formats is illustrated. The data for each TM is treated individually in this pipeline, as slight differences exist between TMs, for example in their effective area \citep{Predehl_2021}.

\subsection{Events}

Events-level eROSITA data is stored in calibrated eventfiles that are created through the eSASS software \citep{Brunner_2022}. They include an eventlist, GTI extensions for all included TMs, and further information, such as the pointing position and live time.

For compatibility with Gammapy slight modifications were made to the eventfiles as follows. 
The \texttt{PI} column containing the event energy was renamed \texttt{ENERGY}.
The time information contained in the \texttt{MJDREF} field in the eventfile header was transferred to fields \texttt{MJDREFI} and \texttt{MJDREFF}.
The deadtime correction is included in the eROSITA eventfiles as a separate extension \texttt{DEADCOR}$n$ (with $n$ the number of the respective TM) with a column \texttt{DEADC}, containing deadtime entries for each $50 \,$ms time interval. As Gammapy takes a single \texttt{DEADC} value, the keyword was filled with the column average.
An \texttt{OBS\_ID} is necessary for the processing of event data, so a substitute \texttt{OBS\_ID} (usually the TM number) was added for each observation.
Additionally the option of adding coordinates of a substitute ``pointing direction", usually the centre of the analysis region, was added for all non-pointed observations to enable the display of event data. Note that this is a fake ``pointing direction'' that does not correspond to the true pointing of the TM. 

\begin{figure}
\resizebox{\hsize}{!}{\includegraphics[clip=true]{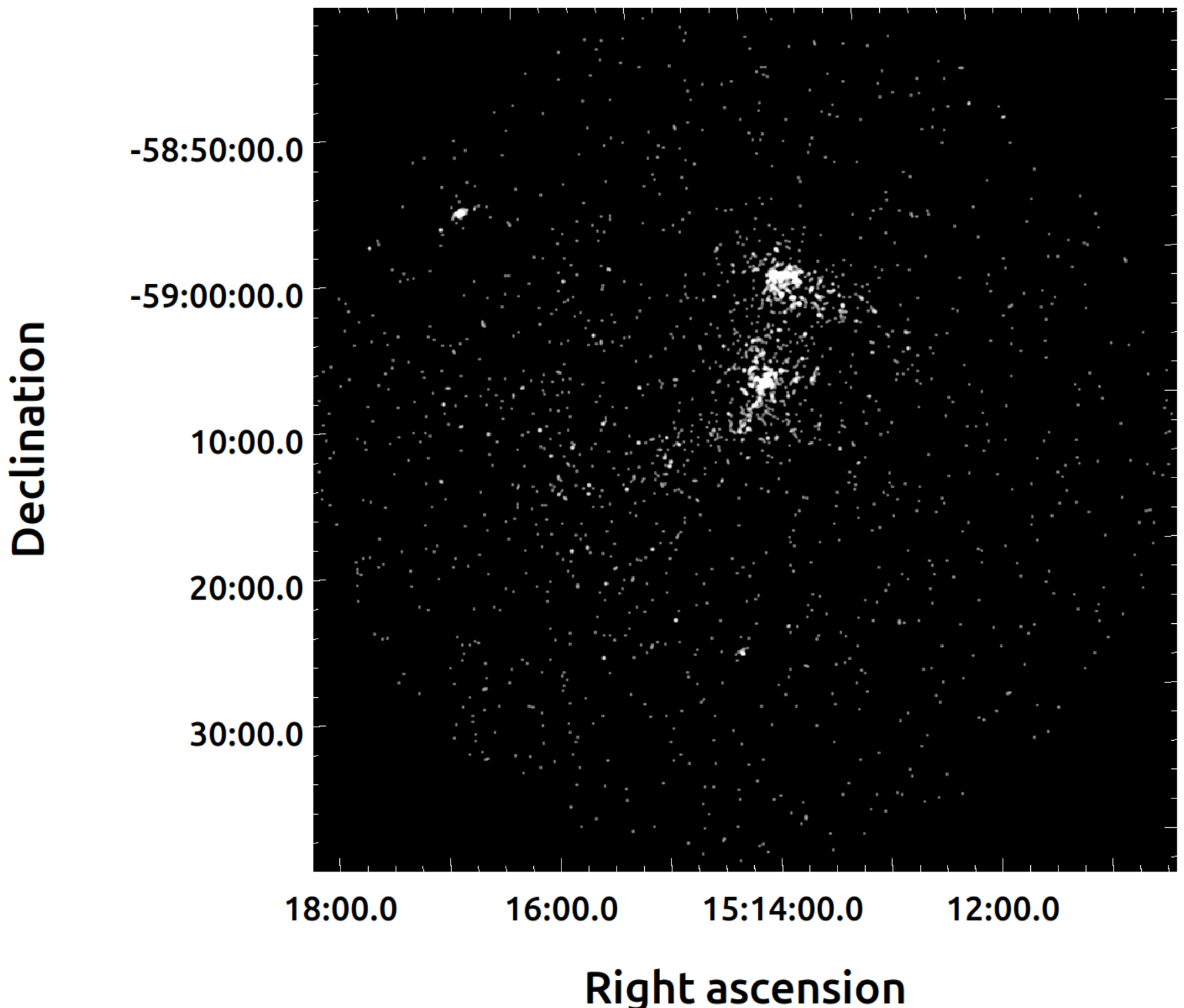}}
\resizebox{\hsize}{!}{\includegraphics[clip=true]{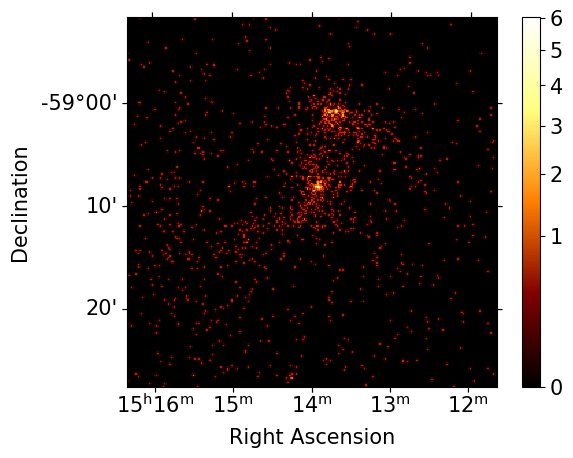}}
\caption{
\footnotesize
eROSITA DR1 TM1 eventfile of MSH 15-52 in ds9 (top) and Gammapy (bottom).}
\label{fig:evtfile}
\end{figure}

With these modifications, the events data could be read into Gammapy via \texttt{EventList.read()}.
A comparison of one of the eventfiles in the image visualisation software ds9 \citep{Joye_2003} and in Gammapy can be seen in Figure \ref{fig:evtfile}.

\subsection{Energy reconstruction}

The energy reconstruction, corresponding to the energy dispersion of eROSITA data is described by the response matrix file (RMF). The RMF is based on calibration measurements and assumed to be only dependent on the pattern filtering of the events. Consequently only one standard RMF is needed for all TMs \citep{Brunner_2022}.

\begin{figure}
\resizebox{\hsize}{!}{\includegraphics[clip=true]{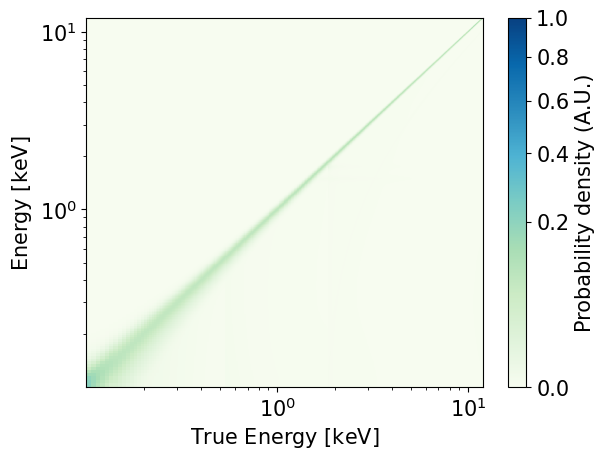}}
\caption{
\footnotesize
eROSITA RMF/energy dispersion matrix plotted in Gammapy.}
\label{fig:rmf}
\end{figure}

The eROSITA RMF is compliant with the OGIP standard\footnote{\url{https://heasarc.gsfc.nasa.gov/docs/heasarc/ofwg/ofwg_recomm.html}}, which is accepted by the Gammapy \texttt{EDispKernel.read()} function\footnote{\url{https://docs.gammapy.org/dev/user-guide/irf/edisp.html}}.
A conversion is, however, presently still needed to read the eROSITA RMF, due to an indexing issue. A solution for this in Gammapy is already being worked on. At present a converter function is used to counteract this problem. Its task is to reduce each value in the \texttt{F\_CHAN} column by exactly one. With this change the RMF can be read by Gammapy. A plot of the eROSITA energy dispersion matrix can be seen in Figure \ref{fig:rmf}.

\subsection{Effective area}
The X-ray effective area is stored in the auxiliary response file (ARF). This is a 1D format (effective area as a function of energy), which is extracted for a particular region of the sky over a particular observation time. Both the ARF and RMF are used in combination with 1D spectra to provide response information.

To obtain a 3D map of the effective area a custom Python script was used. It repeatedly conducts the eSASS \texttt{srctool} task \citep{Brunner_2022} to generate ARFs over small sections of the image field. By moving the selection over the image field a 3D map of the effective area can be created. This map is combined with exposure time information from an exposure time map created with the eSASS task \texttt{expmap} \citep{Brunner_2022} to obtain what is referred to as an exposure map in gamma-ray astronomy. This map can then be handled by Gammapy as a WcsNDMap.



\subsection{Point Spread Function (PSF)}
With a point spread function (PSF) in the arcsec range the resolution of eROSITA is much finer than that of gamma-ray instruments. As such the PSF can be used to account for losses in the extraction of spectra of point sources \citep{Brunner_2022}, but is generally not part of the spectral fits for extended emission. The eROSITA PSF calibration measurements are provided as part of the CALDB in the form of 2D images for different energies and offsets. Radial PSF profiles were extracted from these images and brought into PSF3D format.

Since Gammapy is so far designed for pointing gamma-ray observatories \citep{Donath_2023} (in either RA/DEC or Alt/Az coordinates) creating a PSFMap for the DR1 eROSITA data (taken in SURVEY/SCAN mode) had to be accomplished outside of Gammapy.
Inspired by the approach of \citet{Olivera-Nieto_2022} the PSFMap was created by iterating over all pointing directions of the telescope during the observation. The ratio of time spent in each respective PSF bin was recorded for each pixel and the radial PSF profiles averaged accordingly. Additionally this average was weighted with a factor to account for the decrease in effective area towards the outer edges of the FoV.
The map can then be read by \texttt{PSFMap.read()}.

\subsection{Background}

To estimate the background, Gammapy's MapDatasetOnOff framework was utilised. 
Using a background spectrum extracted from near source free regions of the sky a ``fake'' Off region can be created and automatically subtracted, taking the respective areas and pixel exposures into account.
In the future, a method of fitting the eROSITA background with a multi-component model will be developed. 

\begin{figure*}[t!]
\resizebox{0.5\hsize}{!}{\includegraphics[clip=true]{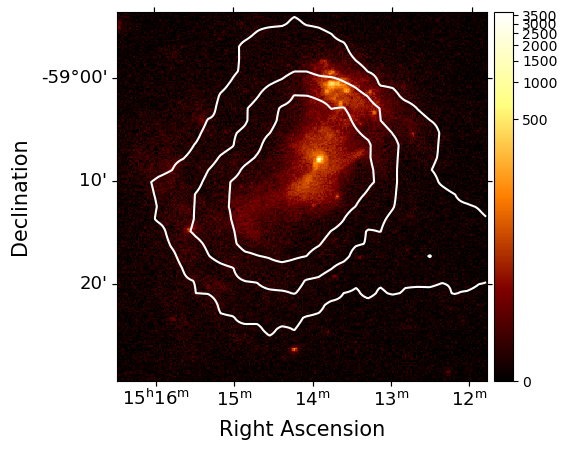}}
\resizebox{0.5\hsize}{!}{\includegraphics[clip=true]{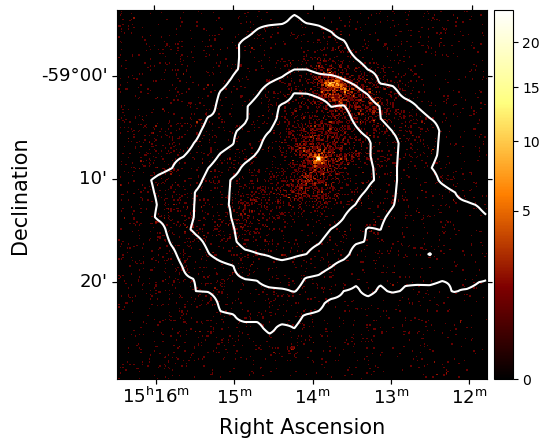}}
\caption{\footnotesize EDR (left) and DR1 (right) eROSITA maps in Gammapy with H.E.S.S. contours (4, 8, and 12 $\sigma$ significance)}
\label{fig:contours_maps}
\end{figure*}

\section{Full maps and comparison to H.E.S.S. data}
The public H.E.S.S. data of MSH 15-52 were read into Gammapy \citep{hess_public_data}, with the accompanying background model fit to the source-free data using FoVBackgroundMaker to create a significance map.
The full maps containing the eROSITA MSH 15-52 data can be seen in Figure \ref{fig:contours_maps} for both the EDR and DR1 data, with H.E.S.S. significance contours at $4$, $8$, and $12 \, \sigma$ overlaid. 

It can be seen that there is very good agreement between not only the location,  but also the elongated shape of the X-ray and gamma-ray sources. This further illustrates the potential for joint multiwavelength analyses utilising the functionalities of Gammapy.

\section{Conclusion and Outlook}
In the course of this work it was shown how eROSITA X-ray data can be adapted into the Gammapy framework on the events-level. The format of the eventfiles, RMFs, ARFs, and PSFs was changed to be compatible with GADF data. We present the first eROSITA counts and IRF maps in Gammapy.

This approach enables large scale studies of pulsar environments at X-ray, as well as at gamma-ray wavelengths, allowing us to investigate properties, such as their morphology, spectra, and to eventually further constrain the characteristics of their underlying particle populations.

A thorough validation of this method as well as a full analysis of eROSITA data in Gammapy is in progress. 



\begin{acknowledgements}
K. Egg and A. M. W. Mitchell are supported by the Deutsche Forschungsgemeinschaft, DFG project number 452934793.

\footnotesize{This work is based on data from eROSITA, the soft X-ray instrument aboard SRG, a joint Russian-German science
mission supported by the Russian Space Agency (Roskosmos), in the interests of the Russian Academy of Sciences
represented by its Space Research Institute (IKI), and the Deutsches Zentrum für Luft- und Raumfahrt (DLR). The
SRG spacecraft was built by Lavochkin Association (NPOL) and its subcontractors, and is operated by NPOL with
support from the Max Planck Institute for Extraterrestrial Physics (MPE). The development and construction of the
eROSITA X-ray instrument was led by MPE, with contributions from the Dr. Karl Remeis Observatory Bamberg \&
ECAP (FAU Erlangen-Nuernberg), the University of Hamburg Observatory, the Leibniz Institute for Astrophysics
Potsdam (AIP), and the Institute for Astronomy and Astrophysics of the University of Tübingen, with the support of
DLR and the Max Planck Society. The Argelander Institute for Astronomy of the University of Bonn and the Ludwig
Maximilians Universität Munich also participated in the science preparation for eROSITA.
The eROSITA data shown here were processed using the eSASS software system developed by the German
eROSITA consortium.}
\end{acknowledgements}

\bibliographystyle{aa}
\bibliography{bibliography.bib}

@ARTICLE{Predehl_2021,
       author = {{Predehl}, P. and {Andritschke}, R. and {Arefiev}, V. and {Babyshkin}, V. and {Batanov}, O. and {Becker}, W. and {B{\"o}hringer}, H. and {Bogomolov}, A. and {Boller}, T. and {Borm}, K. and {Bornemann}, W. and {Br{\"a}uninger}, H. and {Br{\"u}ggen}, M. and {Brunner}, H. and {Brusa}, M. and {Bulbul}, E. and {Buntov}, M. and {Burwitz}, V. and {Burkert}, W. and {Clerc}, N. and {Churazov}, E. and {Coutinho}, D. and {Dauser}, T. and {Dennerl}, K. and {Doroshenko}, V. and {Eder}, J. and {Emberger}, V. and {Eraerds}, T. and {Finoguenov}, A. and {Freyberg}, M. and {Friedrich}, P. and {Friedrich}, S. and {F{\"u}rmetz}, M. and {Georgakakis}, A. and {Gilfanov}, M. and {Granato}, S. and {Grossberger}, C. and {Gueguen}, A. and {Gureev}, P. and {Haberl}, F. and {H{\"a}lker}, O. and {Hartner}, G. and {Hasinger}, G. and {Huber}, H. and {Ji}, L. and {Kienlin}, A. v. and {Kink}, W. and {Korotkov}, F. and {Kreykenbohm}, I. and {Lamer}, G. and {Lomakin}, I. and {Lapshov}, I. and {Liu}, T. and {Maitra}, C. and {Meidinger}, N. and {Menz}, B. and {Merloni}, A. and {Mernik}, T. and {Mican}, B. and {Mohr}, J. and {M{\"u}ller}, S. and {Nandra}, K. and {Nazarov}, V. and {Pacaud}, F. and {Pavlinsky}, M. and {Perinati}, E. and {Pfeffermann}, E. and {Pietschner}, D. and {Ramos-Ceja}, M.~E. and {Rau}, A. and {Reiffers}, J. and {Reiprich}, T.~H. and {Robrade}, J. and {Salvato}, M. and {Sanders}, J. and {Santangelo}, A. and {Sasaki}, M. and {Scheuerle}, H. and {Schmid}, C. and {Schmitt}, J. and {Schwope}, A. and {Shirshakov}, A. and {Steinmetz}, M. and {Stewart}, I. and {Str{\"u}der}, L. and {Sunyaev}, R. and {Tenzer}, C. and {Tiedemann}, L. and {Tr{\"u}mper}, J. and {Voron}, V. and {Weber}, P. and {Wilms}, J. and {Yaroshenko}, V.},
        title = "{The eROSITA X-ray telescope on SRG}",
      journal = {\aap},
         year = 2021,
        month = mar,
       volume = {647},
          eid = {A1},
        pages = {A1},
          doi = {10.1051/0004-6361/202039313},
archivePrefix = {arXiv},
       eprint = {2010.03477},
 primaryClass = {astro-ph.HE},
       adsurl = {https://ui.adsabs.harvard.edu/abs/2021A&A...647A...1P}
}

@ARTICLE{Merloni_2024,
       author = {{Merloni}, A. and {Lamer}, G. and {Liu}, T. and {Ramos-Ceja}, M.~E. and {Brunner}, H. and {Bulbul}, E. and {Dennerl}, K. and {Doroshenko}, V. and {Freyberg}, M.~J. and {Friedrich}, S. and {Gatuzz}, E. and {Georgakakis}, A. and {Haberl}, F. and {Igo}, Z. and {Kreykenbohm}, I. and {Liu}, A. and {Maitra}, C. and {Malyali}, A. and {Mayer}, M.~G.~F. and {Nandra}, K. and {Predehl}, P. and {Robrade}, J. and {Salvato}, M. and {Sanders}, J.~S. and {Stewart}, I. and {Tub{\'\i}n-Arenas}, D. and {Weber}, P. and {Wilms}, J. and {Arcodia}, R. and {Artis}, E. and {Aschersleben}, J. and {Avakyan}, A. and {Aydar}, C. and {Bahar}, Y.~E. and {Balzer}, F. and {Becker}, W. and {Berger}, K. and {Boller}, T. and {Bornemann}, W. and {Br{\"u}ggen}, M. and {Brusa}, M. and {Buchner}, J. and {Burwitz}, V. and {Camilloni}, F. and {Clerc}, N. and {Comparat}, J. and {Coutinho}, D. and {Czesla}, S. and {Dannhauer}, S.~M. and {Dauner}, L. and {Dauser}, T. and {Dietl}, J. and {Dolag}, K. and {Dwelly}, T. and {Egg}, K. and {Ehl}, E. and {Freund}, S. and {Friedrich}, P. and {Gaida}, R. and {Garrel}, C. and {Ghirardini}, V. and {Gokus}, A. and {Gr{\"u}nwald}, G. and {Grandis}, S. and {Grotova}, I. and {Gruen}, D. and {Gueguen}, A. and {H{\"a}mmerich}, S. and {Hamaus}, N. and {Hasinger}, G. and {Haubner}, K. and {Homan}, D. and {Ider Chitham}, J. and {Joseph}, W.~M. and {Joyce}, A. and {K{\"o}nig}, O. and {Kaltenbrunner}, D.~M. and {Khokhriakova}, A. and {Kink}, W. and {Kirsch}, C. and {Kluge}, M. and {Knies}, J. and {Krippendorf}, S. and {Krumpe}, M. and {Kurpas}, J. and {Li}, P. and {Liu}, Z. and {Locatelli}, N. and {Lorenz}, M. and {M{\"u}ller}, S. and {Magaudda}, E. and {Mannes}, C. and {McCall}, H. and {Meidinger}, N. and {Michailidis}, M. and {Migkas}, K. and {Mu{\~n}oz-Giraldo}, D. and {Musiimenta}, B. and {Nguyen-Dang}, N.~T. and {Ni}, Q. and {Olechowska}, A. and {Ota}, N. and {Pacaud}, F. and {Pasini}, T. and {Perinati}, E. and {Pires}, A.~M. and {Pommranz}, C. and {Ponti}, G. and {Poppenhaeger}, K. and {P{\"u}hlhofer}, G. and {Rau}, A. and {Reh}, M. and {Reiprich}, T.~H. and {Roster}, W. and {Saeedi}, S. and {Santangelo}, A. and {Sasaki}, M. and {Schmitt}, J. and {Schneider}, P.~C. and {Schrabback}, T. and {Schuster}, N. and {Schwope}, A. and {Seppi}, R. and {Serim}, M.~M. and {Shreeram}, S. and {Sokolova-Lapa}, E. and {Starck}, H. and {Stelzer}, B. and {Stierhof}, J. and {Suleimanov}, V. and {Tenzer}, C. and {Traulsen}, I. and {Tr{\"u}mper}, J. and {Tsuge}, K. and {Urrutia}, T. and {Veronica}, A. and {Waddell}, S.~G.~H. and {Willer}, R. and {Wolf}, J. and {Yeung}, M.~C.~H. and {Zainab}, A. and {Zangrandi}, F. and {Zhang}, X. and {Zhang}, Y. and {Zheng}, X.},
        title = "{The SRG/eROSITA all-sky survey. First X-ray catalogues and data release of the western Galactic hemisphere}",
      journal = {\aap},
         year = 2024,
        month = feb,
       volume = {682},
          eid = {A34},
        pages = {A34},
          doi = {10.1051/0004-6361/202347165},
archivePrefix = {arXiv},
       eprint = {2401.17274},
 primaryClass = {astro-ph.HE},
       adsurl = {https://ui.adsabs.harvard.edu/abs/2024A&A...682A..34M}
}

@ARTICLE{Donath_2023,
       author = {{Donath}, Axel and {Terrier}, R{\'e}gis and {Remy}, Quentin and {Sinha}, Atreyee and {Nigro}, Cosimo and {Pintore}, Fabio and {Kh{\'e}lifi}, Bruno and {Olivera-Nieto}, Laura and {Ruiz}, Jose Enrique and {Br{\"u}gge}, Kai and {Linhoff}, Maximilian and {Contreras}, Jose Luis and {Acero}, Fabio and {Aguasca-Cabot}, Arnau and {Berge}, David and {Bhattacharjee}, Pooja and {Buchner}, Johannes and {Boisson}, Catherine and {Carreto Fidalgo}, David and {Chen}, Andrew and {de Bony de Lavergne}, Mathieu and {de Miranda Cardoso}, Jos{\'e} Vinicius and {Deil}, Christoph and {F{\"u}{\ss}ling}, Matthias and {Funk}, Stefan and {Giunti}, Luca and {Hinton}, Jim and {Jouvin}, L{\'e}a and {King}, Johannes and {Lefaucheur}, Julien and {Lemoine-Goumard}, Marianne and {Lenain}, Jean-Philippe and {L{\'o}pez-Coto}, Rub{\'e}n and {Mohrmann}, Lars and {Morcuende}, Daniel and {Panny}, Sebastian and {Regeard}, Maxime and {Saha}, Lab and {Siejkowski}, Hubert and {Siemiginowska}, Aneta and {Sip{\H{o}}cz}, Brigitta M. and {Unbehaun}, Tim and {van Eldik}, Christopher and {Vuillaume}, Thomas and {Zanin}, Roberta},
        title = "{Gammapy: A Python package for gamma-ray astronomy}",
      journal = {\aap},
         year = 2023,
        month = oct,
       volume = {678},
          eid = {A157},
        pages = {A157},
          doi = {10.1051/0004-6361/202346488},
archivePrefix = {arXiv},
       eprint = {2308.13584},
 primaryClass = {astro-ph.IM},
       adsurl = {https://ui.adsabs.harvard.edu/abs/2023A&A...678A.157D}
}

@misc{hess_public_data,
  doi = {10.5281/ZENODO.1421098},
  
  url = {https://zenodo.org/record/1421098},
  
  author = {{H.E.S.S. Collaboration}
},
  
  language = {en},
  
  title = {H.E.S.S. First Public Test Data Release},
  
  publisher = {Zenodo},
  
  year = {2018},
  
  copyright = {Open Access}
}

@ARTICLE{Brunner_2022,
       author = {{Brunner}, H. and {Liu}, T. and {Lamer}, G. and {Georgakakis}, A. and {Merloni}, A. and {Brusa}, M. and {Bulbul}, E. and {Dennerl}, K. and {Friedrich}, S. and {Liu}, A. and {Maitra}, C. and {Nandra}, K. and {Ramos-Ceja}, M.~E. and {Sanders}, J.~S. and {Stewart}, I.~M. and {Boller}, T. and {Buchner}, J. and {Clerc}, N. and {Comparat}, J. and {Dwelly}, T. and {Eckert}, D. and {Finoguenov}, A. and {Freyberg}, M. and {Ghirardini}, V. and {Gueguen}, A. and {Haberl}, F. and {Kreykenbohm}, I. and {Krumpe}, M. and {Osterhage}, S. and {Pacaud}, F. and {Predehl}, P. and {Reiprich}, T.~H. and {Robrade}, J. and {Salvato}, M. and {Santangelo}, A. and {Schrabback}, T. and {Schwope}, A. and {Wilms}, J.},
        title = "{The eROSITA Final Equatorial Depth Survey (eFEDS). X-ray catalogue}",
      journal = {\aap},
         year = 2022,
        month = may,
       volume = {661},
          eid = {A1},
        pages = {A1},
          doi = {10.1051/0004-6361/202141266},
archivePrefix = {arXiv},
       eprint = {2106.14517},
 primaryClass = {astro-ph.HE},
       adsurl = {https://ui.adsabs.harvard.edu/abs/2022A&A...661A...1B}
}

@ARTICLE{Mohrmann_2019,
       author = {{Mohrmann}, L. and {Specovius}, A. and {Tiziani}, D. and {Funk}, S. and {Malyshev}, D. and {Nakashima}, K. and {van Eldik}, C.},
        title = "{Validation of open-source science tools and background model construction in {\ensuremath{\gamma}}-ray astronomy}",
      journal = {\aap},
         year = 2019,
        month = dec,
       volume = {632},
          eid = {A72},
        pages = {A72},
          doi = {10.1051/0004-6361/201936452},
archivePrefix = {arXiv},
       eprint = {1910.08088},
 primaryClass = {astro-ph.IM},
       adsurl = {https://ui.adsabs.harvard.edu/abs/2019A&A...632A..72M}
}

@INPROCEEDINGS{Joye_2003,
       author = {{Joye}, W.~A. and {Mandel}, E.},
        title = "{New Features of SAOImage DS9}",
    booktitle = {ADASS XII},
         year = 2003,
       series = {Astron. Soc. of the Pacific Conference Series},
       volume = {295},
        month = jan,
        pages = {489},
       adsurl = {https://ui.adsabs.harvard.edu/abs/2003ASPC..295..489J}
}

@ARTICLE{Giunti_2022,
       author = {{Giunti}, L. and {Acero}, F. and {Kh{\'e}lifi}, B. and {Kosack}, K. and {Lemi{\`e}re}, A. and {Terrier}, R.},
        title = "{Constraining leptonic emission scenarios for the PeVatron candidate HESS J1702{\ensuremath{-}}420 with deep XMM-Newton observations}",
      journal = {\aap},
         year = 2022,
        month = nov,
       volume = {667},
          eid = {A130},
        pages = {A130},
          doi = {10.1051/0004-6361/202244696},
archivePrefix = {arXiv},
       eprint = {2209.09566},
 primaryClass = {astro-ph.HE},
       adsurl = {https://ui.adsabs.harvard.edu/abs/2022A&A...667A.130G}
}

@ARTICLE{Nigro_2019,
       author = {{Nigro}, C. and {Deil}, C. and {Zanin}, R. and {Hassan}, T. and {King}, J. and {Ruiz}, J.~E. and {Saha}, L. and {Terrier}, R. and {Br{\"u}gge}, K. and {N{\"o}the}, M. and {Bird}, R. and {Lin}, T.~T.~Y. and {Aleksi{\'c}}, J. and {Boisson}, C. and {Contreras}, J.~L. and {Donath}, A. and {Jouvin}, L. and {Kelley-Hoskins}, N. and {Khelifi}, B. and {Kosack}, K. and {Rico}, J. and {Sinha}, A.},
        title = "{Towards open and reproducible multi-instrument analysis in gamma-ray astronomy}",
      journal = {\aap},
         year = 2019,
        month = may,
       volume = {625},
          eid = {A10},
        pages = {A10},
          doi = {10.1051/0004-6361/201834938},
archivePrefix = {arXiv},
       eprint = {1903.06621},
 primaryClass = {astro-ph.HE},
       adsurl = {https://ui.adsabs.harvard.edu/abs/2019A&A...625A..10N}
}

@misc{Rosillo_2024,
      title={A Unified Multi-Wavelength Data Analysis Workflow with gammapy}, 
      author={M. Nievas Rosillo and F. Acero and J. Otero-Santos and M. Vazquez Acosta and R. Terrier and A. Arbet-Engels},
      year={2024},
      eprint={2409.20487},
      archivePrefix={arXiv},
      primaryClass={astro-ph.HE},
      url={https://arxiv.org/abs/2409.20487}, 
}

@ARTICLE{Gaensler_2006,
       author = {{Gaensler}, Bryan M. and {Slane}, Patrick O.},
        title = "{The Evolution and Structure of Pulsar Wind Nebulae}",
      journal = {\araa},
         year = 2006,
        month = sep,
       volume = {44},
       number = {1},
        pages = {17-47},
          doi = {10.1146/annurev.astro.44.051905.092528},
archivePrefix = {arXiv},
       eprint = {astro-ph/0601081},
 primaryClass = {astro-ph},
       adsurl = {https://ui.adsabs.harvard.edu/abs/2006ARA&A..44...17G}
}

@INCOLLECTION{Mitchell_2022,
       author = {{Mitchell}, A.~M.~W. and {Gelfand}, J.},
        title = "{Pulsar Wind Nebulae}",
    booktitle = {Handbook of X-ray and Gamma-ray Astrophysics},
         year = 2022,
       editor = {{Bambi}, Cosimo and {Sangangelo}, Andrea},
          eid = {61},
        pages = {61},
          doi = {10.1007/978-981-16-4544-0_157-1},
       adsurl = {https://ui.adsabs.harvard.edu/abs/2022hxga.book...61M}
}

\end{document}